\documentclass[%
 reprint,
nofootinbib,
 amsmath,amssymb,
 aps,
]{revtex4-2}

\usepackage{graphicx}
\usepackage{dcolumn}
\usepackage{bm}
\usepackage{lipsum}

\usepackage{tabularx}
\newcolumntype{Y}{>{\centering\arraybackslash}X}

\usepackage{amsmath}
\usepackage{amsfonts}
\usepackage{amssymb}
\usepackage{graphicx, rotating}
\usepackage{epstopdf}
\usepackage{epsfig}
\usepackage{latexsym}
\usepackage{multirow}
\usepackage{color}
\usepackage{slashed}
\usepackage{xcolor}
\usepackage{float}
\usepackage{mathtools}
\usepackage[colorlinks=true,linkcolor=red]{hyperref}

\newcommand{\ba}{\begin{eqnarray}}
\newcommand{\ea}{\end{eqnarray}}

\begin{document}

\title{
\LARGE{
When CP requires $\bar\theta=0$, not $\bar\theta=\pi$}
}

\author{Luca Vecchi}
 \affiliation{Istituto Nazionale di Fisica Nucleare (INFN), Sezione di Padova, Italy}
 \email{luca.vecchi@pd.infn.it}


\begin{abstract}

Imposing CP or P forces the QCD topological angle to be either $\bar\theta=0$ or $\bar\theta=\pi$. However, only the former is phenomenologically viable. This implies that the assumption of CP or P alone cannot provide a framework for unambiguously solving the Strong CP problem. We show that $\bar\theta=0$ is naturally selected when the assumption of CP is combined with the hypothesis that the Standard Model is embedded in a suitable gauge group.

\end{abstract}

\maketitle


\section{A hidden ``quality problem"}

Solutions to the Strong CP Problem based on an exact CP or P symmetry \cite{Nelson:1983zb,Barr:1984qx,Babu:1989rb,Barr:1991qx,Kuchimanchi:1995rp,Hiller:2001qg,Valenti:2021xjp,Feruglio:2023uof,Hall:2024xbd} offer an appealing alternative to the popular QCD axion.\footnote{See the appendix.} 
In that approach the absence of CP violation in the strong interactions is recast as a model-building challenge: CP or P must be spontaneously broken in a way that reproduces the Standard Model --- including the CKM matrix --- while ensuring that all corrections to the QCD theta angle remain below approximately $10^{-10}$. Achieving this goal requires scenarios with highly non-generic structures and minimal sources of CP and flavor violation beyond the Standard Model. An optimist might view this as a compelling sign that the Strong CP Problem serves as a valuable guide to the correct ultraviolet completion of the Standard Model.

If I have to identify the main qualitative advantage of these approaches over the QCD axion, I would say that --- unlike the axion --- they are based on an {\emph{exact symmetry principle}}. As a result, the quantum field theorist does not need to know the complete UV theory to assess the validity of the solution: the mere existence of a CP (or P) symmetry is sufficient. This makes a conventional quantum field theory (QFT) framework structurally resilient to any potential ``quality problem". 

By contrast, the QCD axion solution relies on the existence of an anomalous symmetry of high quality. And since anomalous symmetries neither walk nor quack like true symmetries, a careful model-builder naturally feels compelled to question the soundness of the assumption. That is, to establish whether the axion truly solves the Strong CP Problem the model-builder would like to convince him or herself that a theory with an anomalous symmetry of high quality can actually be constructed. However, he or she would also quickly realize that providing a concrete answer is difficult without a theory of gravity, and that perhaps a more suitable setup for approaching the axion ``quality problem" might be offered by string theory. Yet, our current understanding of the vacuum of quantum gravity does not allow for definitive conclusions. It is thus fair to say that, at present, the axion quality problem cannot be reliably quantified.

Constructions based on exact CP (or P) have their own challenges, but fortunately, a potential CP (or P) quality problem is not among them. Of course, assessing the {\emph{plausibility}} of the hypothesis of an exact CP (or P) does require some knowledge of the UV theory, and a truly reliable judgment will likely have to await a deeper understanding of quantum gravity. 
\footnote{Concretely, it has been shown that CP (or P) can be gauge symmetries in certain extra-dimensional gravitational theories, including string theory \cite{Strominger:1985it,Choi:1992xp,Dine:1992ya}. That makes the basic assumption more attractive. Yet, in order to address the Strong CP Problem such a symmetry should remain unbroken when compactifying to 4-dimensions, a feature that may not be as generic as one may hope \cite{Strominger:1985it,Dine:2015jga}.} Still, it is a significant relief that we do not need to address such a formidable task in order to claim that exact CP (or P) provides a solution to the Strong CP Problem. From the perspective of a modest low-energy theorist, this is undeniably a practical advantage.

And yet, there is a subtlety that appears to have been entirely overlooked in the literature. The hypothesis of CP alone is compatible with {\emph{two}} distinct possibilities for the rescaling-invariant theta angle:
\ba\label{theta0}
\bar\theta=0~~~~~{\text{or}}~~~~~
\bar\theta=\pi. 
\ea
However, as emphasized in \cite{Crewther:1979pi}, experimental evidence indicates that real-world QCD has $\bar\theta\approx0$. A hypothetical QCD with $\bar\theta=\pi$ would still yield a vanishing neutron electric dipole moment, but it would fail to reproduce the observed meson spectrum and is thus firmly excluded.\footnote{QCD-like theories actually break CP spontaneously at $\bar\theta=\pi$ for particular choices of the quark masses, as first pointed out by Dashen (see also \cite{Witten:1980sp,Creutz:1995wf,Smilga:1998dh} and \cite{Gaiotto:2017yup,Gaiotto:2017tne}).}

The realization that the sole hypothesis of CP (or P) is not sufficient to guarantee $\bar\theta=0$ may represent a dramatic blow for theories that rely on these symmetries. As argued above, their main advantage over the QCD axion lies in the apparent ability to remain agnostic about the UV completion. However this advantage seems lost if the ambiguity \eqref{theta0} cannot be resolved within an effective field theory framework. A scenario that ``potentially works fifty percent of the times" does not constitute a genuine solution unless one can meaningfully quantify its likelihood of success --- something that unavoidably requires knowledge of the UV theory. In this sense, the dichotomy between $\bar\theta=0$ and $\bar\theta=\pi$ reveals an intrinsic ``quality problem" for models based on spontaneous CP (or P) violation. In their competitor, the QCD axion, such ambiguity is completely avoided since $\bar\theta=0$ is dynamically favored over $\bar\theta=\pi$ \cite{Vafa:1984xg}.

As far I can tell, among the models proposed so far the only ones that resolve the ambiguity \eqref{theta0} are the grand-unified scenarios mentioned at the end of \cite{Barr:1991qx}. In those constructions, P is combined with a mirror symmetry acting on the entire Standard Model gauge group. When QCD emerges as the diagonal subgroup one finds $\bar\theta=0$ regardless of the value of the topological angles of the parent groups (see also \cite{Bonnefoy:2023afx}). However, the vast majority of models based on P do not involve such a doubling of QCD, and constructions based on CP do not include it either. This raises an obvious question: is there a general IR mechanism to resolve \eqref{theta0} that applies to these more conventional frameworks?

Fortunately, as I will show below, there exists an elegant and robust mechanism to unambiguously select the option $\bar\theta=0$ in general CP (or P) invariant effective field theories. This does not require introducing new IR degrees of freedom, but rather hinges on a careful choice of the underlying gauge symmetry.

\section{Engineering $\bar\theta=0$}

Let us start from the beginning. Viewed as a spurion, the theta angle transforms under P and CP as
\ba\label{PonTheta1}
\bar\theta\to-\bar\theta.
\ea
Therefore, only theories in which $\bar\theta$ is physically equivalent to $-\bar\theta$ can be CP (or P) invariant. Because $\bar\theta$ in QCD has period $2\pi$, the condition $\bar\theta\sim-\bar\theta$ can be satisfied for both values in \eqref{theta0}. Consequently, an exact CP symmetry can suppress CP violation in the strong interactions but cannot distinguish between $\bar\theta=0,\pi$. For example, a fermion mass matrix is forced to be hermitian by P, or real by CP, but these symmetries alone do not fix the sign of its determinant.

The ambiguity \eqref{theta0} is inevitable {\emph{if $\bar\theta$ has period $2\pi$.} Yet, suppose that $\bar\theta$ has instead a larger period of $2\pi p$ for some number $p$. In this exotic scenario, CP (or P) would be conserved if there exists an integer $n$ such that $\bar\theta=-\bar\theta+2\pi p n\in[0,2\pi p)$, that is if
\ba\label{Nperiod}
\bar\theta=0~~~~~{\text{or}}~~~~~
\bar\theta=\pi p. 
\ea
We thus discover that in a gauge theory in which $p$ is an even integer, CP (or P) would force $\bar\theta$ to be a multiple of $2\pi$. The ambiguity in Eq.~\eqref{theta0} can therefore be resolved by embedding QCD in a gauge group with even $p$. Our task is to see how this can be concretely achieved.

Recall that the periodicity of the topological angle of a gauge theory may be identified inspecting the topological charges carried by its instantons. In an $SU(N)$ gauge theory instantons have integer charges and consequently $\bar\theta$ has period $2\pi$. In contrast, given a $SU(N)/{\mathbb Z}_N$ gauge theory on a non-trivial manifold, where ${\mathbb Z}_N$ is the center subgroup, there exist fractionally charged instantons of topological charge $1/N$, and the theta angle acquires an enlarged periodicity of $2\pi N$ \cite{tHooft:1979rtg,tHooft:1981nnx}. This enlarged periodicity can also be diagnosed by examining the spectrum of line operators in the theory \cite{Aharony:2013hda}. In particular, under the shift $\bar\theta\to\bar\theta+2\pi$ the charges of the line operators of the $SU(N)/{\mathbb Z}_N$ theory transform non-trivially via the Witten effect, and only $\bar\theta\to\bar\theta+2\pi N$ leaves the full spectrum invariant.

Interestingly, the Standard Model particle content allows for a gauge group with non-trivial quotient $\Gamma={\mathbb Z}_2,{\mathbb Z}_3,{\mathbb Z}_6$ and correspondingly a non-trivial spectrum of line operators. However, it turns out that only the periodicity of the topological angle associated to the $U(1)$ hypercharge factor is affected by these discrete quotients, while that of QCD retains its usual $2\pi$ periodicity \cite{Tong:2017oea}. I will discuss the reason behind this below. For now, it is important to note that, evidently, to enlarge the periodicity of the QCD $\bar\theta$ angle one must embed the Standard Model into a larger group.

Finding a unified gauge group with the desired property requires some work. A natural direction might be to consider embedding the Standard Model into an $SU(2N)/{\mathbb Z}_p$. However, this requires that all matter fields transform in representations that are neutral under ${\mathbb Z}_p$, i.e. with $N$-ality multiple of $p$. For instance, a unified theory based on $SU(2N)/{\mathbb Z}_2$ demands that all chiral matter resides in representations with even number of indices ($2$, $4$, etc.). I was not able to construct an anomaly-free theory of this kind, in which all exotic states are vector-like under the Standard Model subgroup.

We might have a better chance of success if we consider groups of the form  
\ba\label{G}
G=\frac{G_1\times G_2\times \cdots}{{\mathbb Z}_2},
\ea
where for simplicity I assume that the center of all factors $G_{a=1,2,\cdots}$ contains a ${\mathbb Z}_2$ subgroup, and that the quotient in \eqref{G} corresponds to the diagonal of these ${\mathbb Z}_2$'s. Groups like \eqref{G} are obviously not the most general candidates that can meet our requirements, but they represent a promising subset to explore. In particular, the presence of more factors $G_a$ significantly broadens the range of representations that matter fields can carry. With that setup it becomes possible for matter to transform under a representation of $G_a$ that is non-singlet under the quotient $\mathbb Z_2$, provided it is simultaneously charged under another $\mathbb Z_2$-odd representation of a different factor $G_b$, in a manner analogous to what occurs in the Standard Model$/\Gamma$. However, a crucial subtlety arises when considering product groups of the form \eqref{G}. It is precisely this subtlety that prevents the theta angles associated to the non-abelian gauge groups in $SU(3)\times SU(2)\times U(1)/\Gamma$ from exhibiting an enlarged periodicity.

Denote by $\bar\theta_a$ the physical theta angle associated with each factor $G_a$. In a given topological sector the theta angles enter the path integral through terms like $\exp{i{\bar\theta}_a Q_a}$ with $\vec Q=(Q_1,Q_2,\cdots)$ the topological charges. As well-known, ordinary BPST instantons \cite{Belavin:1975fg} can be constructed individually for each $G_a$'s, and carry integer charges $n_a\in{\mathbb Z}$. Hence each $\bar\theta_a$ must have a period that is an integer multiple of $2\pi$. However, the fractional instantons that may arise for $G_a/{\mathbb Z}_2$ {\emph{cannot}} be constructed independently. Because the quotient ${\mathbb Z}_2$ acts on all $G_a$ factors, the characteristic class that obstructs lifting a $G_a/{\mathbb Z}_2$-bundle to its simply connected cover is shared accross all factors. Relatedly, on a manifold of non-trivial topology, the transition functions  transform non-trivially under the very same ${\mathbb Z}_2$. As a result, the fractional instantons associated with each $G_a/{\mathbb Z}_2$ are linked together. Denoting by $q_a$ the smallest non-trivial topological charge associated to $G_a/{\mathbb Z}_2$, then: 
\ba\label{Q}
\vec Q=\vec n+m\,\vec q,
\ea
where $m\in{\mathbb Z}$. For reference, Table \ref{tab1} collects the minimal instanton charges for a variety of groups, taken from \cite{Aharony:2013hda,Cordova:2019uob,Apruzzi:2021vcu}.\footnote{In the literature $m$ is often expressed as $\int_{\cal M}{\cal P}(w_2)/2$, where ${\cal M}$ is the space-time manifold and ${\cal P}(w_2)$ is the Pontryagin square of the second Stiefel-Whitney class $w_2$ of the principal $G$-bundle. Compared to \cite{Aharony:2013hda,Cordova:2019uob,Apruzzi:2021vcu}, Eq. \eqref{Q} already includes the factor of $2$ that comes from the fact that $\int_{\cal M}{\cal P}(w_2)$ is always even on spin manifolds.}

\begin{center}
\begin{table}[t]
\begin{tabular}{c|c|c}
Group $G_a$ & $~~q~~$ & ${\mathbb Z}_2\subset {\text{Center}}(G_a)$\\
  \hline\hline
   $SO(4N+2)$  & $1/4$ & Obvious
 \\\hline
  $SU(4N+2)$  & $1/2$ & Obvious
  \\
 $Spin(8N+4)$  & $1/2$ & Either ${\mathbb Z}^s_2$ or ${\mathbb Z}^c_2$
 \\
    $SO(4N+4)$  & $1/2$ & Obvious
\\
$USp(4N+2)$ & $1/2$ & Obvious
\\\hline
$SU(4N)$ & $1$ & Obvious
\\
$Spin(4+N)$ &  $1$ & Obvious or ${\mathbb Z}^{s+c}_2$
\\
$USp(4N)$ & $1$ & Obvious
\end{tabular}
\caption{List of the minimal topological charge of $G_a/\mathbb Z_2$ for various $G_a$, with $N\in{\mathbb Z}_{\geq1}$. The quotient in \eqref{G} is embedded in the center of $G_a$ in the obvious way; the only non-trivial cases are for $Spin(4N)$, whose center is ${\mathbb Z}^{s}_2\times {\mathbb Z}^{c}_2$. For simplicity we did not consider abelian groups. Furthermore, $Spin(4)$ and $SO(4)=Spin(4)/{\mathbb Z}_2$ are not shown; they admit fractional instantons of charge $1/2$ and have two topological angles.}
\label{tab1}
\end{table}
\end{center}

\vspace{-1.cm}

The general form of \eqref{Q} implies that an integer shift 
\ba
\Delta\vec{\bar\theta}=2\pi\vec k 
\ea
of the theta angles of \eqref{G}, for some $\vec k\in{\mathbb Z}\times {\mathbb Z}\times\cdots$, leaves the theory invariant when $\vec k\cdot\vec q~{\in}~{\mathbb Z}$. This fact explains, for example, why in the case of $G=SU(2)\times SU(2)/{\mathbb Z}_2$ the pairs $(\bar\theta_1,\bar\theta_2)$ and $(\bar\theta_1+2\pi,\bar\theta_2+2\pi)$ are completely equivalent despite $SU(2)/{\mathbb Z}_2$ admits instantons with fractional charge $1/2$. Indeed, in such a theory $\vec q=(1/2,1/2)$ and one has the identification $(\bar\theta_1,\bar\theta_2)\sim(\bar\theta_1+4\pi,\bar\theta_2)\sim(\bar\theta_1+2\pi,\bar\theta_2+2\pi)$ while $(\bar\theta_1,\bar\theta_2)$ and $(\bar\theta_1+2\pi,\bar\theta_2)$ correspond to two inequivalent theories.

The subtlety we face is now clear. Suppose we embed the Standard Model in a group like \eqref{G}, with QCD fully contained in $G_1$. Whenever $\vec k\cdot\vec q~{\in}~{\mathbb Z}$, a shift $\bar\theta_1\to\bar\theta_1+2\pi$ can be compensated by shifts in the theta angles of the other groups, even though $G_1/{\mathbb Z}_2$ would, in its own, admit fractional instantons. Hence, to forbid the identification $\bar\theta_1\sim\bar\theta_1+2\pi$ and thus realize my program, the gauge group \eqref{G} must be chosen so that the charge vector $\vec q$ satisfies
\ba\label{condition}
\vec k\cdot\vec q~\slashed{\in}~{\mathbb Z}~~~\forall~~~\vec{k}=(1,k_2,k_3\cdots),
\ea
with $k_{a\neq1}\in{\mathbb Z}$. 

While the formula \eqref{condition} was derived for the specific structure \eqref{G}, the conclusion is completely general. For instance, the fractional instantons in an $SU(N)\times U(1)/\mathbb Z_N$ theory necessarily involve both the $SU(N)$ and $U(1)$ factors, see \cite{Anber:2021upc} for an explicit construction following the seminal work by 't Hooft. Consequently, a $2\pi$ shift of the theta angle of the non-abelian factor can always be compensated by a corresponding shift of the $U(1)$ theta angle. This is why the presence of a non-trivial quotient $\Gamma$ in the Standard Model does not lead to an enlarged periodicity of the theta angles associated with either $SU(3)$ or $SU(2)$ \cite{Tong:2017oea}.

It is easy to see that none of the grand-unified groups that have been proposed so far satisfies my requirement \eqref{condition}. The most popular ones, $SU(5)$ and $Spin(10)$\footnote{In the phenomenological literature, this is commonly referred to as ``$SO(10)$ grand unification". Strictly speaking, though, the gauge group in that scenario is $Spin(10)$. 
} with matter in the $\overline{\bf 5}\oplus{\bf 10}$ and ${\bf 16}$ respectively, do not have fractional instantons, and so the theta angles in both theories are periodic of $2\pi$. Consider next the following incarnation $SU(4)\times SU(2)\times SU(2)/{{\mathbb Z}_2}$ of the Pati-Salam group \cite{Pati:1974yy}, with matter transforming as $({\bf 4},{\bf 2},{\bf 1})\oplus(\overline{\bf 4},{\bf 1},{\bf 2})$. This group is of the form anticipated in \eqref{G}, with QCD embedded into the $SU(4)$ factor. Unfortunately, $SU(4)/{\mathbb Z}_2=SO(6)$ does not admit fractional instantons whereas $SU(2)/{\mathbb Z}_2$ does. So $\vec q=(1,1/2,1/2)$ and \eqref{condition} cannot hold. Trinification, with an $SU(3)\times SU(3)\times SU(3)/{\mathbb Z}_3$ gauge group and the standard fermion embedding $({\bf 3},{\bf 3},{\bf 1})\oplus(\overline{\bf 3},{\bf 1},\overline{\bf 3})\oplus({\bf 1},\overline{\bf 3},{\bf 3})$, has $\vec q=(1/3,1/3,1/3)$. Furthermore, $SU(6)\times SU(2)/{\mathbb Z}_2$, with matter transforming as $(\overline{\bf 6},{\bf 2})\oplus({\bf 15},{\bf 1})$ \cite{Dimopoulos:1985xs}, has $\vec q=(1/2,1/2)$. Neither of them satisfies \eqref{condition}.

With the charges listed in Table \ref{tab1}, Eq. \eqref{condition} reduces to the requirement that the effective period of the theta angle of $G_1$ (the group that contains QCD by assumption) must be larger than the others, i.e. ${1}/{q_1}>{1}/{q_{a\neq1}}$. There are two possible solutions. The first is to choose $G_1$ so that $G_1/{\mathbb Z}_2$ admits, in isolation, fractional instantons of charge $q_1=1/4$, while all the other factors $G_{a\neq1}/{\mathbb Z}_2$'s in \eqref{G} are chosen among those with $q_{a\neq1}=1/2$ or $0$. The second possibility is to take $q_1=1/2$ and $q_{a\neq1}=0$. 

The remaining challenge is to find a $G$ in which we can embed the Standard Model in such a way that all exotic matter remains vector-like under the $SU(3)\times SU(2)\times U(1)$ subgroup. While this can be achieved, the construction of a concrete model is not straightforward. As a proof of concept, here is a simple illustrative example:
\ba\label{GGG}
\frac{SU(6)\times USp(4)\times USp(4)}{{\mathbb Z}_2},
\ea
with fermions transforming as $({\bf 6},{\bf 4},{\bf 1})\oplus(\overline{\bf 6},{\bf 1},{\bf 4})$. To see how the Standard Model emerges from \eqref{GGG}, first assume that both $USp(4)$ factors are broken down to an $SU(2)$ subgroup with the branching ${\bf 4}={\bf 2}\oplus{\bf 1}\oplus{\bf 1}$; this can be achieved through scalars in the ${\bf 5}\oplus{\bf 10}$. The fermionic content then decomposes into a chiral sector $({\bf 6},{\bf 2},{\bf 1})\oplus(\overline{\bf 6},{\bf 1},{\bf 2})\in SU(6)\times SU(2)\times SU(2)/{\mathbb Z}_2$ plus two copies of vector-like states $({\bf 6},{\bf 1},{\bf 1})\oplus(\overline{\bf 6},{\bf 1},{\bf 1})$, which can acquire large masses. The $SU(6)$ is instead broken down to $SU(4)\times SU(2)'$ with the branching ${\bf 6}=({\bf 4},{\bf 1})\oplus({\bf 1},{\bf 2})$ induced by the vacuum expectation value of a scalar in the ${\bf 15}$. At this stage, the structure corresponds to a Pati-Salam model times an additional $SU(2)'$. All fermions charged under the extra $SU(2)'$ are exotic and vector-like, leaving the chiral fermion content identical to that of Pati-Salam. At low energies the Standard Model gauge group emerges as the familiar $SU(3)\times SU(2)\times U(1)/{\mathbb Z}_6$. Optionally, the model \eqref{GGG} may be endowed with a left-right symmetry exchanging the two $USp(4)$'s, as required to implement parity. 

The key feature of the model is that the charge vector is $\vec q=(1/2,1,1)$. The spectrum of line operators compatible with \eqref{GGG} is thus affected by a shift of $2\pi$ of the theta angle of $SU(6)$ but not by arbitrary $2\pi k$ shifts of the theta angles of the $USp(4)$ factors. Only after a $4\pi$ rotation of the former one is back to the original theory. Therefore, imposing CP in such a theory forces $\bar\theta_1$ to be $0$ or $2\pi$, and $\bar\theta_2,\bar\theta_3$ to be $0$ or $\pi$. This is precisely what we were looking for.

Now, regardless of how CP is spontaneously broken, the QCD topological angle obtained by matching the Standard Model to \eqref{GGG} will coincide with that of the parent $G_1$, up to radiative corrections that must be small in any realistic solution to the Strong CP Problem, i.e. 
\ba\label{key}
\bar\theta\approx\bar\theta_1. 
\ea
Therefore, successful theories of spontaneous CP (or P) breaking based on the gauge group \eqref{GGG} will automatically ensure that the real QCD topological angle is close to either $0$ or $2\pi$. And since these two values are practically indistinguishable from the point of view of low-energy QCD, the strong CP phase will effectively be indistinguishable from zero, in agreement with experimental observations.

It is important to understand how \eqref{key} is reconciled with the statement that the theta angle in standard QCD is $2\pi$-periodic. More generally, it is very instructive to see how the mechanism we propose appears from the perspective of a low-energy observer who sees only the Standard Model gauge group. The key point is that the actual periodicity of the topological angle is determined by the global structure of the underlying gauge group, and is not manifest within the effective field theory. In fact, the UV theory based on $G$ includes not only ordinary particles, such as quarks and gluons, but also colored topological defects, such as Georgi-Glashow monopoles with masses proportional to the scale where $G\to SU(3)\times SU(2)\times U(1)/\Gamma$. Although the renormalizable version of QCD is invariant under $\bar\theta\to\bar\theta+2\pi$, the full spectrum of the topological defects is only invariant under $\bar\theta\to\bar\theta+2\pi p$. Any experiment probing QCD at low energies would thus find that the physics at $\bar\theta=0$ and $\bar\theta= 2\pi$ is the same up to minuscule corrections due to the virtual exchange of topological defects. 

I could have simply assumed the existence of a spectrum of Standard Model--charged defects with the desired property. However, that would not have addressed the ``quality problem" as I introduced it. It was therefore necessary to provide a genuinely quantum field-theoretic explanation.

\section{Outlook}

The sole hypothesis that CP or P is spontaneously broken --- while keeping radiative corrections to $\bar\theta$ under control --- is not sufficient to explain why the QCD topological angle is close to $\bar\theta=0$ rather than to the experimentally excluded CP-conserving value $\bar\theta=\pi$. To resolve this hidden ``quality problem" and unambiguously address the Strong CP Problem via CP (or P), one can embed the Standard Model into a larger gauge group $G$ in such a way that the topological angle of QCD has an extended periodicity of $2\pi p$, with $p\geq2$ an even integer.

I identified a broad class of such gauge groups $G$, see \eqref{G} and condition \eqref{condition}, and briefly discussed an explicit embedding. Different group structures can be studied along similar lines. I am confident that a systematic exploration of the class \eqref{G}, as well as its generalizations, will uncover more elegant and compelling models than the one presented here.


The upshot is that robust QFT solutions to the Strong CP problem based on spontaneous CP (or P) violation can be constructed, provided the model-builder incorporates a new requirement: the underlying gauge symmetry must belong to the class identified here, or to a suitable generalization. This requirement has two important implications. First, because some form of unification is unavoidable, the flavor structure of the theory cannot be a carbon copy of that of the Standard Model; one must therefore ensure that no large radiative corrections to $\bar\theta$ are induced. Second, in addition to the CP-breaking sector, the theory requires a scalar sector to break the unified group down to the Standard Model; this breaking must avoid introducing dangerous CP-odd phases. Both issues are intrinsically model-dependent and cannot be analyzed in generality. They deserve detailed study but do not pose fundamental obstacles to building compelling scenarios.

In conclusion, it is quite suggestive that some form of grand unification may be necessary to solve the Strong CP Problem via a fundamental CP (or P) symmetry. An optimist might view this as yet another hint that we may be on the right track.

\acknowledgements

I am deeply indebted to F. Apruzzi for explaining the origin of \eqref{Q} and for clarifying the results of \cite{Aharony:2013hda}. I would also like to thank J. Davighi for insightful discussions and valuable comments on the manuscript, and to F. Feruglio and A. Valenti for helping me sharpen my thoughts on \cite{Kaplan:2025bgy}.  

This work was partly supported by the Italian MIUR under contract 202289JEW4 (Flavors: dark and intense), the Iniziativa Specifica “Physics at the Energy, Intensity, and Astroparticle Frontiers” (APINE) of INFN, the European Union’s Horizon 2020 research and innovation programme under the Marie Sklodowska-Curie grant agreement n. 860881- HIDDeN and n. 101086085 – ASYMMETRY, and the European Union - Next Generation EU through the MUR PRIN2022 Grant n.202289JEW4.

\appendix

\section*{A comment on \cite{Kaplan:2025bgy}}
\label{app1}

Given that this work concerns the use of CP (or P) in the context of the strong CP problem, I should comment on the recent paper \cite{Kaplan:2025bgy}. 

In Yang--Mills theories, the $\theta$-parameter can be introduced either as a coefficient in the Hamiltonian $H$ or {\emph{equivalently}} as a boundary condition associated with large gauge transformations \cite{Jackiw:1979ur}.\footnote{See also the beautiful lectures by D. Tong \cite{Tong}.}
 In the latter case, and working in temporal gauge $A_0 = 0$, the boundary condition reads:
\begin{equation}
\label{BCs}
U | \Psi \rangle = e^{-i\theta} | \Psi \rangle.
\end{equation}
Here $|\Psi \rangle$ is an arbitrary state of the Hilbert space, and $U$ is the unitary operator implementing large gauge transformations, defined by its action on configuration states as $U |A_i(x)\rangle = |A^g_i(x)\rangle$, where $A^g_i=gA_ig^\dagger-ig\partial_ig^\dagger$ is the gauge-transformed field for a transformation $g(x)$ with unit winding number.

Regardless of whether it appears as an operator coefficient or via \eqref{BCs}, $\theta$ affects the physics and is therefore subject to symmetry constraints. In particular, if the theory is invariant under parity, both the Hamiltonian and \eqref{BCs} must be compatible with this symmetry.

Let $P$ denote the unitary operator implementing parity as $P |A_i(x)\rangle = |{-A}_i(x_P)\rangle$. In a P-symmetric theory, the part of the algebra relevant to us is:
\begin{equation}
\label{algebra}
[H, P] = 0, \quad [H, U] = 0, \quad P U P^\dagger = U^\dagger.
\end{equation}
The last relation follows from an explicit computation in the basis $|A_i(x)\rangle$, using the fact that $g(x_P)=g^\dagger(x)$, which reflects the P-odd nature of the topological charge.

Now, in a generic quantum mechanical system, the algebra \eqref{algebra} implies that $H$, $P$, and $U$ cannot be simultaneously diagonalized. This is precisely the situation in the toy example discussed in \cite{Kaplan:2025bgy}, where $U$ represents discrete translations in a periodic potential. There, energy eigenstates are generically superpositions of parity eigenstates with different Bloch momenta, or superpositions of $U$-eigenstates with different parity.

However, gauge theories are qualitatively different \cite{Jackiw:1979ur}. If large gauge transformations are to remain unbroken, then the vacuum must satisfy \eqref{BCs}, and since the entire physical Hilbert space is obtained acting on the vacuum with gauge-invariant operators, all states must satisfy \eqref{BCs}. In other words, the operator $U$ defines a superselection sector: the Hilbert space is restricted to states with a fixed eigenvalue of $U$. Within such a sector, $U = e^{-i\theta}$ acts as a multiple of the identity. 

In a general P-invariant theory, both $|\Psi\rangle$ and $P^\dagger|\Psi\rangle$ belong to the same Hilbert space, and we just saw that in a {\emph{gauge}} theory both must have the same $U$-eigenvalue $e^{-i\theta}$. However, the last relation in \eqref{algebra} says that $P^\dagger|\Psi\rangle$ has eigenvalue $e^{+i\theta}$. As a result, in gauge theories the algebra \eqref{algebra} can only be satisfied if $e^{-i\theta}=e^{+i\theta}$, as also emphasized in \cite{Kuchimanchi:2025pqb}: parity invariance requires that $\theta$ in \eqref{BCs} be either $0$ or $\pi$.

Because both \eqref{algebra} and the boundary condition~\eqref{BCs} are part of the \emph{definition} of the gauge theory, they are unaffected by spontaneous symmetry breaking. If the theory undergoes spontaneous parity violation, the Hilbert space within a given $\theta$-sector splits further into disjoint subspaces, exchanged by the action of $P$. Nevertheless, the value of $\theta$ in \eqref{BCs} must remain the {same} --- either $0$ or $\pi$ --- across all these disjoint subspaces. Otherwise, the breaking of P would be explicit.

\bibliography{biblio}

\begin{thebibliography}{10}

\bibitem{Nelson:1983zb}
Ann~E. Nelson.
\newblock {Naturally Weak CP Violation}.
\newblock {\em Phys. Lett. B}, 136:387--391, 1984.
\newblock \href {https://doi.org/10.1016/0370-2693(84)92025-2}
  {\path{doi:10.1016/0370-2693(84)92025-2}}.

\bibitem{Barr:1984qx}
Stephen~M. Barr.
\newblock {Solving the Strong CP Problem Without the Peccei-Quinn Symmetry}.
\newblock {\em Phys. Rev. Lett.}, 53:329, 1984.
\newblock \href {https://doi.org/10.1103/PhysRevLett.53.329}
  {\path{doi:10.1103/PhysRevLett.53.329}}.

\bibitem{Babu:1989rb}
K.~S. Babu and Rabindra~N. Mohapatra.
\newblock {A Solution to the Strong {CP} Problem Without an Axion}.
\newblock {\em Phys. Rev. D}, 41:1286, 1990.
\newblock \href {https://doi.org/10.1103/PhysRevD.41.1286}
  {\path{doi:10.1103/PhysRevD.41.1286}}.

\bibitem{Barr:1991qx}
Stephen~M. Barr, D.~Chang, and G.~Senjanovic.
\newblock {Strong CP problem and parity}.
\newblock {\em Phys. Rev. Lett.}, 67:2765--2768, 1991.
\newblock \href {https://doi.org/10.1103/PhysRevLett.67.2765}
  {\path{doi:10.1103/PhysRevLett.67.2765}}.

\bibitem{Kuchimanchi:1995rp}
Ravi Kuchimanchi.
\newblock {Solution to the strong CP problem: Supersymmetry with parity}.
\newblock {\em Phys. Rev. Lett.}, 76:3486--3489, 1996.
\newblock \href {http://arxiv.org/abs/hep-ph/9511376}
  {\path{arXiv:hep-ph/9511376}}, \href
  {https://doi.org/10.1103/PhysRevLett.76.3486}
  {\path{doi:10.1103/PhysRevLett.76.3486}}.

\bibitem{Hiller:2001qg}
Gudrun Hiller and Martin Schmaltz.
\newblock {Solving the Strong CP Problem with Supersymmetry}.
\newblock {\em Phys. Lett. B}, 514:263--268, 2001.
\newblock \href {http://arxiv.org/abs/hep-ph/0105254}
  {\path{arXiv:hep-ph/0105254}}, \href
  {https://doi.org/10.1016/S0370-2693(01)00814-0}
  {\path{doi:10.1016/S0370-2693(01)00814-0}}.

\bibitem{Valenti:2021xjp}
Alessandro Valenti and Luca Vecchi.
\newblock {Super-soft CP violation}.
\newblock {\em JHEP}, 07(152):152, 2021.
\newblock \href {http://arxiv.org/abs/2106.09108} {\path{arXiv:2106.09108}},
  \href {https://doi.org/10.1007/JHEP07(2021)152}
  {\path{doi:10.1007/JHEP07(2021)152}}.

\bibitem{Feruglio:2023uof}
Ferruccio Feruglio, Alessandro Strumia, and Arsenii Titov.
\newblock {Modular invariance and the QCD angle}.
\newblock {\em JHEP}, 07:027, 2023.
\newblock \href {http://arxiv.org/abs/2305.08908} {\path{arXiv:2305.08908}},
  \href {https://doi.org/10.1007/JHEP07(2023)027}
  {\path{doi:10.1007/JHEP07(2023)027}}.

\bibitem{Hall:2024xbd}
Lawrence Hall, Claudio~Andrea Manzari, and Bea Noether.
\newblock {Strong CP and flavor in multi-Higgs theories}.
\newblock {\em Phys. Rev. D}, 111(11):115012, 2025.
\newblock \href {http://arxiv.org/abs/2407.14585} {\path{arXiv:2407.14585}},
  \href {https://doi.org/10.1103/rxml-5gd2} {\path{doi:10.1103/rxml-5gd2}}.

\bibitem{Strominger:1985it}
A.~Strominger and Edward Witten.
\newblock {New Manifolds for Superstring Compactification}.
\newblock {\em Commun. Math. Phys.}, 101:341, 1985.
\newblock \href {https://doi.org/10.1007/BF01216094}
  {\path{doi:10.1007/BF01216094}}.

\bibitem{Choi:1992xp}
Ki-woon Choi, David~B. Kaplan, and Ann~E. Nelson.
\newblock {Is CP a gauge symmetry?}
\newblock {\em Nucl. Phys. B}, 391:515--530, 1993.
\newblock \href {http://arxiv.org/abs/hep-ph/9205202}
  {\path{arXiv:hep-ph/9205202}}, \href
  {https://doi.org/10.1016/0550-3213(93)90082-Z}
  {\path{doi:10.1016/0550-3213(93)90082-Z}}.

\bibitem{Dine:1992ya}
Michael Dine, Robert~G. Leigh, and Douglas~A. MacIntire.
\newblock {Of CP and other gauge symmetries in string theory}.
\newblock {\em Phys. Rev. Lett.}, 69:2030--2032, 1992.
\newblock \href {http://arxiv.org/abs/hep-th/9205011}
  {\path{arXiv:hep-th/9205011}}, \href
  {https://doi.org/10.1103/PhysRevLett.69.2030}
  {\path{doi:10.1103/PhysRevLett.69.2030}}.

\bibitem{Dine:2015jga}
Michael Dine and Patrick Draper.
\newblock {Challenges for the Nelson-Barr Mechanism}.
\newblock {\em JHEP}, 08:132, 2015.
\newblock \href {http://arxiv.org/abs/1506.05433} {\path{arXiv:1506.05433}},
  \href {https://doi.org/10.1007/JHEP08(2015)132}
  {\path{doi:10.1007/JHEP08(2015)132}}.

\bibitem{Crewther:1979pi}
R.~J. Crewther, P.~Di~Vecchia, G.~Veneziano, and Edward Witten.
\newblock {Chiral Estimate of the Electric Dipole Moment of the Neutron in
  Quantum Chromodynamics}.
\newblock {\em Phys. Lett. B}, 88:123, 1979.
\newblock [Erratum: Phys.Lett.B 91, 487 (1980)].
\newblock \href {https://doi.org/10.1016/0370-2693(79)90128-X}
  {\path{doi:10.1016/0370-2693(79)90128-X}}.

\bibitem{Witten:1980sp}
Edward Witten.
\newblock {Large N Chiral Dynamics}.
\newblock {\em Annals Phys.}, 128:363, 1980.
\newblock \href {https://doi.org/10.1016/0003-4916(80)90325-5}
  {\path{doi:10.1016/0003-4916(80)90325-5}}.

\bibitem{Creutz:1995wf}
Michael Creutz.
\newblock {Quark masses and chiral symmetry}.
\newblock {\em Phys. Rev. D}, 52:2951--2959, 1995.
\newblock \href {http://arxiv.org/abs/hep-th/9505112}
  {\path{arXiv:hep-th/9505112}}, \href
  {https://doi.org/10.1103/PhysRevD.52.2951}
  {\path{doi:10.1103/PhysRevD.52.2951}}.

\bibitem{Smilga:1998dh}
Andrei~V. Smilga.
\newblock {QCD at theta similar to pi}.
\newblock {\em Phys. Rev. D}, 59:114021, 1999.
\newblock \href {http://arxiv.org/abs/hep-ph/9805214}
  {\path{arXiv:hep-ph/9805214}}, \href
  {https://doi.org/10.1103/PhysRevD.59.114021}
  {\path{doi:10.1103/PhysRevD.59.114021}}.

\bibitem{Gaiotto:2017yup}
Davide Gaiotto, Anton Kapustin, Zohar Komargodski, and Nathan Seiberg.
\newblock {Theta, Time Reversal, and Temperature}.
\newblock {\em JHEP}, 05:091, 2017.
\newblock \href {http://arxiv.org/abs/1703.00501} {\path{arXiv:1703.00501}},
  \href {https://doi.org/10.1007/JHEP05(2017)091}
  {\path{doi:10.1007/JHEP05(2017)091}}.

\bibitem{Gaiotto:2017tne}
Davide Gaiotto, Zohar Komargodski, and Nathan Seiberg.
\newblock {Time-reversal breaking in QCD$_{4}$, walls, and dualities in 2 + 1
  dimensions}.
\newblock {\em JHEP}, 01:110, 2018.
\newblock \href {http://arxiv.org/abs/1708.06806} {\path{arXiv:1708.06806}},
  \href {https://doi.org/10.1007/JHEP01(2018)110}
  {\path{doi:10.1007/JHEP01(2018)110}}.

\bibitem{Vafa:1984xg}
Cumrun Vafa and Edward Witten.
\newblock {Parity Conservation in QCD}.
\newblock {\em Phys. Rev. Lett.}, 53:535, 1984.
\newblock \href {https://doi.org/10.1103/PhysRevLett.53.535}
  {\path{doi:10.1103/PhysRevLett.53.535}}.

\bibitem{Bonnefoy:2023afx}
Quentin Bonnefoy, Lawrence Hall, Claudio~Andrea Manzari, and Christiane Scherb.
\newblock {Colorful Mirror Solution to the Strong CP Problem}.
\newblock {\em Phys. Rev. Lett.}, 131(22):221802, 2023.
\newblock \href {http://arxiv.org/abs/2303.06156} {\path{arXiv:2303.06156}},
  \href {https://doi.org/10.1103/PhysRevLett.131.221802}
  {\path{doi:10.1103/PhysRevLett.131.221802}}.

\bibitem{tHooft:1979rtg}
Gerard 't~Hooft.
\newblock {A Property of Electric and Magnetic Flux in Nonabelian Gauge
  Theories}.
\newblock {\em Nucl. Phys. B}, 153:141--160, 1979.
\newblock \href {https://doi.org/10.1016/0550-3213(79)90595-9}
  {\path{doi:10.1016/0550-3213(79)90595-9}}.

\bibitem{tHooft:1981nnx}
Gerard 't~Hooft.
\newblock {Some Twisted Selfdual Solutions for the Yang-Mills Equations on a
  Hypertorus}.
\newblock {\em Commun. Math. Phys.}, 81:267--275, 1981.
\newblock \href {https://doi.org/10.1007/BF01208900}
  {\path{doi:10.1007/BF01208900}}.

\bibitem{Aharony:2013hda}
Ofer Aharony, Nathan Seiberg, and Yuji Tachikawa.
\newblock {Reading between the lines of four-dimensional gauge theories}.
\newblock {\em JHEP}, 08:115, 2013.
\newblock \href {http://arxiv.org/abs/1305.0318} {\path{arXiv:1305.0318}},
  \href {https://doi.org/10.1007/JHEP08(2013)115}
  {\path{doi:10.1007/JHEP08(2013)115}}.

\bibitem{Tong:2017oea}
David Tong.
\newblock {Line Operators in the Standard Model}.
\newblock {\em JHEP}, 07:104, 2017.
\newblock \href {http://arxiv.org/abs/1705.01853} {\path{arXiv:1705.01853}},
  \href {https://doi.org/10.1007/JHEP07(2017)104}
  {\path{doi:10.1007/JHEP07(2017)104}}.

\bibitem{Belavin:1975fg}
A.~A. Belavin, Alexander~M. Polyakov, A.~S. Schwartz, and Yu.~S. Tyupkin.
\newblock {Pseudoparticle Solutions of the Yang-Mills Equations}.
\newblock {\em Phys. Lett. B}, 59:85--87, 1975.
\newblock \href {https://doi.org/10.1016/0370-2693(75)90163-X}
  {\path{doi:10.1016/0370-2693(75)90163-X}}.

\bibitem{Cordova:2019uob}
Clay C{\'o}rdova, Daniel~S. Freed, Ho~Tat Lam, and Nathan Seiberg.
\newblock {Anomalies in the Space of Coupling Constants and Their Dynamical
  Applications II}.
\newblock {\em SciPost Phys.}, 8(1):002, 2020.
\newblock \href {http://arxiv.org/abs/1905.13361} {\path{arXiv:1905.13361}},
  \href {https://doi.org/10.21468/SciPostPhys.8.1.002}
  {\path{doi:10.21468/SciPostPhys.8.1.002}}.

\bibitem{Apruzzi:2021vcu}
Fabio Apruzzi, Sakura Schafer-Nameki, Lakshya Bhardwaj, and Jihwan Oh.
\newblock {The Global Form of Flavor Symmetries and 2-Group Symmetries in 5d
  SCFTs}.
\newblock {\em SciPost Phys.}, 13(2):024, 2022.
\newblock \href {http://arxiv.org/abs/2105.08724} {\path{arXiv:2105.08724}},
  \href {https://doi.org/10.21468/SciPostPhys.13.2.024}
  {\path{doi:10.21468/SciPostPhys.13.2.024}}.

\bibitem{Anber:2021upc}
Mohamed~M. Anber and Erich Poppitz.
\newblock {Nonperturbative effects in the Standard Model with gauged 1-form
  symmetry}.
\newblock {\em JHEP}, 12:055, 2021.
\newblock \href {http://arxiv.org/abs/2110.02981} {\path{arXiv:2110.02981}},
  \href {https://doi.org/10.1007/JHEP12(2021)055}
  {\path{doi:10.1007/JHEP12(2021)055}}.

\bibitem{Pati:1974yy}
Jogesh~C. Pati and Abdus Salam.
\newblock {Lepton Number as the Fourth Color}.
\newblock {\em Phys. Rev. D}, 10:275--289, 1974.
\newblock [Erratum: Phys.Rev.D 11, 703--703 (1975)].
\newblock \href {https://doi.org/10.1103/PhysRevD.10.275}
  {\path{doi:10.1103/PhysRevD.10.275}}.

\bibitem{Dimopoulos:1985xs}
S.~Dimopoulos and Lawrence~J. Hall.
\newblock {FLIPPING AWAY PROTON DECAY}.
\newblock {\em Nucl. Phys. B}, 255:633--647, 1985.
\newblock \href {https://doi.org/10.1016/0550-3213(85)90157-9}
  {\path{doi:10.1016/0550-3213(85)90157-9}}.

\bibitem{Kaplan:2025bgy}
David~E. Kaplan, Tom Melia, and Surjeet Rajendran.
\newblock {What can solve the Strong CP problem?}
\newblock 5 2025.
\newblock \href {http://arxiv.org/abs/2505.08358} {\path{arXiv:2505.08358}}.

\bibitem{Jackiw:1979ur}
R.~Jackiw.
\newblock {Introduction to the Yang-Mills Quantum Theory}.
\newblock {\em Rev. Mod. Phys.}, 52:661--673, 1980.
\newblock \href {https://doi.org/10.1103/RevModPhys.52.661}
  {\path{doi:10.1103/RevModPhys.52.661}}.

\bibitem{Tong}
David Tong.
\newblock {Lectures on Gauge Theory
  \url{https://www.damtp.cam.ac.uk/user/tong/gaugetheory.html}}.

\bibitem{Kuchimanchi:2025pqb}
Ravi Kuchimanchi.
\newblock {Parity solves the Strong CP problem}.
\newblock 6 2025.
\newblock \href {http://arxiv.org/abs/2506.01911} {\path{arXiv:2506.01911}}.

\end{thebibliography}

\bibliographystyle{unsrturl}


\end{document}